\documentclass[aps,prl,twocolumn,superscriptaddress]{revtex4}
\usepackage[dvips]{graphics}

\begin{document}


\title{The turbulent mean-velocity profile: it is all in the spectrum 
}


\author{Gustavo Gioia}
\affiliation{Department of Mechanical Science and Engineering, University of Illinois, Urbana, IL 61801}
\author{Nicholas Guttenberg}
\affiliation{Department of Physics, University of Illinois, Urbana, IL 61801}
\author{Nigel Goldenfeld}
\affiliation{Department of Physics,  University of Illinois, Urbana, IL 61801}
\author{Pinaki Chakraborty}
\affiliation{Department of Geology, University of Illinois, Urbana, IL 61801}


\date{\today}

\begin{abstract}

It has long been surmised that the mean-velocity profile (MVP) 
of a pipe flow is closely related 
to the spectrum of turbulent energy. Here we perform a 
spectral analysis to identify the eddies that 
dominate the production of shear stress via 
momentum transfer. This analysis allows us to express 
the MVP as a functional of the spectrum. Each part of 
the MVP relates to a specific spectral range: 
the buffer layer to the dissipative range, the log layer to 
the inertial range, and the wake to 
the energetic range. The parameters of the spectrum set the 
thickness of the viscous layer, the 
amplitude of the buffer layer, and the amplitude of the wake. 



\end{abstract}


\maketitle


 Although most flows 
   in nature and technology
  are turbulent flows over confining walls, 
  these flows have remained amongst 
 the least understood phenomena of 
 classical physics \cite{noncapisco}.
 Take the simplest example (an example
 whose applications are legion):
 the turbulent flow in a long cylindrical 
 pipe with a 
 cross section of radius $R$
 and a smooth  internal wall. If the flux is kept 
  steady, the velocity of the flow at a distance
 $y$ from the wall of the pipe 
 may be averaged over a long period 
 of time to obtain a  mean velocity $u$.
 The function $u(y)$ is called the
 mean-velocity profile (MVP), and there
 is a MVP for each value of the Reynolds 
 number, ${\rm Re}\equiv UR/\nu$,  where $U$ is the 
 mean velocity of the flow (i.e., 
 the flux divided by the cross-sectional 
 area of the pipe) 
 and $\nu$ is the kinematic viscosity
of the fluid \cite{tl,prandtl}. (Re quantifies 
 the relative importance of inertia and 
  viscosity in the flow; the higher the Re, the more
 turbulent the flow.)
  MVPs were first measured 
 70 years ago \cite{niku} and
  have recently been the subject of 
 exacting experiments \cite{exper} and numerous computational
  simulations \cite{computgen,comput}. Theory has meanwhile
 lagged well behind experiments
 and simulations. 

\begin{figure} 
\centering
\resizebox{3in}{!}{\includegraphics{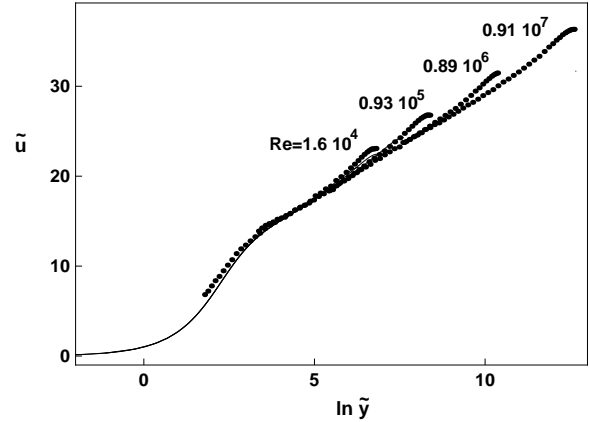}}
\caption{\label{expe} Log-linear plots of the 
 MVPs in the ``wall variables'' ${\tilde u}$ and ${\tilde y}$
  for four values of Re. The symbols are from experiments 
 \cite{exper} and the solid lines from simulations \cite{comput}. Each  MVP
 extends from the wall (which corresponds always to ${\tilde y}=0$)
 to the centerline of the pipe (which 
 corresponds to a value of ${\tilde y}$ that depends on 
 Re); the MVPs collapse on a single MVP close to 
 the wall. Focusing on any one MVP, we can parse the horizontal
 axis from left to right to find
  the ``viscous layer'' (where the MVP has a positive curvature),
 the ``buffer layer'' (where the MVP has a noticeable
  negative curvature),
 the ``log layer'' (where the MVP follows the log law of 
 the wall (\ref{llw})), and a ``wake'' (where the
  MVP overshoots the log law of the wall
 close to the centerline of the pipe) \cite{layers}. }
\end{figure}

The first theory came soon after the
 earliest experiments. 
Ludwig Prandtl
 showed that when plotted in terms of the 
 dimensionless 
 ``wall variables'' ${\tilde u}$ and ${\tilde y}$,
 the MVPs for different values of 
 Re collapse into a single MVP close to the wall
  (Fig.~\ref{expe}) \cite{prandtl}.
 The wall variables are 
  defined by
 ${\tilde u}\equiv u/\sqrt{\tau_0/\rho}$ and
  ${\tilde y}\equiv (\sqrt{\tau_0/\rho})/\nu$, 
 where $\rho$ is the density of the fluid and 
 $\tau_0$ is the shear stress---or shear force per unit 
 area---that develops between the flow and the 
  wall of the pipe.
 Prandlt also showed that over most of its domain
 ${\tilde u}({\tilde y})$
 follows the ``log law of the wall,''
 \begin{equation} \label{llw}
{\tilde u}({\tilde y})=
{{1}\over{\kappa}} \ln {\tilde y} +B,  
\end{equation}
 where $\kappa$ (the ``K\'arm\'an constant'') and $B$ are 
 dimensionless constants which can be estimated
 by fitting experimental data \cite{prandtl}.
 
 Numerous variants of the log law of the wall have
 been proposed since the time of Prandtl, and recently it has 
 been argued that the MVPs for different values of Re might not
 be logarithmic after all. Instead, they might be 
 power laws with Re-dependent exponents and a 
 logarithmic envelope \cite{baren}.
 Yet the original theory of Prandtl and 
 this recent theory of Barenblatt have 
  been predicated on dimensional analysis and 
 similarity assumptions (they differ in the 
 similarity assumptions \cite{baren}), without reference
 to the turbulent spectrum. As a result, 
 these theories 
cannot be used to relate, for example, the
 log law of the wall or the constants thereof
 to the turbulent spectrum. 
(The turbulent 
spectrum may be said to embody the fabric of 
 a turbulent state; in practice it is an account 
 of the form in which the kinetic 
energy is apportioned among the eddies of different sizes 
in the flow.)  
  Our aim here is to find
  the  missing link between the MVP and 
 the turbulent spectrum.


 We adopt the
 phenomenological imagery of 
 ``turbulent eddies'' \cite{rich,frisch}
 and use the 
 spectrum of turbulent energy \cite{tl,pope} at the wavenumber $k$,
 $E(k)$, to determine the velocity of 
 the eddies of size $s$, $v_s$, in the form
$v_s^2=\int_{1/s}^\infty E(k) dk$, where 
$E(k)={{2}\over{3}}\,(\kappa_\varepsilon \varepsilon)^{2/3}\, 
 k^{-5/3}\,c_d(\eta k)\,c_e(R k)$.
  Here $\kappa_\varepsilon$ is a dimensionless parameter,
   $\varepsilon$ is the turbulent power per unit mass \cite{taylor},
 $\eta=\nu^{3/4}\,\varepsilon^{-1/4}$ is the 
 viscous lengthscale \cite{tl},
  $R$ is the largest lengthscale in the flow, 
 ${{2}\over{3}}\,(\kappa_\varepsilon \varepsilon)^{2/3}\, k^{-5/3}$ is
 the Kolmog\'orov spectrum \cite{kolmo,aniso}, 
 and $c_d$ and $c_e$ 
 are dimensionless correction functions---the 
 dissipative-range correction and the
 energetic-range correction, respectively.
 For the dissipative-range correction 
   we adopt the usual exponential
 form, $c_d(\eta k)=\exp(-\beta_d\eta k)$, and for 
 the energetic-range correction 
 the form proposed by  K\'arm\'an, 
 $c_e(R k)=(1+ \beta_e(R k)^{-2})^{-17/6}$,
 where $\beta_d$ and $\beta_e$ are nonnegative 
  dimensionless parameters \cite{pope}.

By introducing the 
  dimensionless variable $\xi\equiv s k$, we 
 can write $v_s= (\kappa_\varepsilon \varepsilon s)^{1/3}\sqrt{I}$,
 where  $I={\mathcal I}(\eta/s,s/R)\equiv {{2}\over{3}}\int_{1}^\infty 
 \xi^{-5/3} \exp(-\xi\,\beta_d\,\eta/s)\, 
 (1+ \beta_e (s/R)^2/\xi^2)^{-17/6}d\xi$.
 For $s$ in the inertial range ($\eta\ll s \ll R$), 
  $I=1$, and therefore 
$v_s= (\kappa_\varepsilon \varepsilon s)^{1/3}$
 \cite{tl,frisch,pope}, with the implication that
   the velocity of an eddy of the inertial range 
   (i.e., an eddy of size $\eta\ll s\ll R$)
  increases with 
 the size of the eddy.
 The same result, $v_s= (\kappa_\varepsilon \varepsilon s)^{1/3}$,
 may  be derived  from Kolmog\'orov's four-fifth law, which 
  also gives the estimate $\kappa_\varepsilon=4/5$ \cite{ff}.
  For $s$ outside of the inertial range, 
  $I<1$. 
  It follows that  
 an eddy of the dissipative range or the energetic range
  (i.e., an eddy of size $s\approx \eta$ or $s\approx R$,
 respectively) has a velocity
 $v_s< (\kappa_\varepsilon \varepsilon s)^{1/3}$---i.e.,
   the eddy is {\it slower\/} than an imaginary eddy of the same 
 size in the inertial range.

 Consider now the flow in a smooth-walled pipe. 
  The energy equation gives and expression for $\varepsilon$
in the form $\varepsilon=\tau_t u'/\rho$, where 
 $\tau_t$ is the turbulent shear stress,
 $u'\equiv du/dy$, and $y$ is the distance to the wall
 \cite{pope}.
The sum of the turbulent 
 shear stress $\tau_t$ and the viscous shear stress 
 $\rho \nu u'$ is the total shear stress $\tau$, 
 which in a pipe of radius $R$ 
  is given by 
 $\tau=\tau_0 (1-y/R)$, where $\tau_0$ is the 
  shear stress at the wall \cite{pope}. Using the definition of the 
 friction factor, $f\equiv \tau_0/\rho U^2$, we can write
 $\tau_0=\rho f U^2$ \cite{pope}; it follows that
 \begin{equation} \label{tau}
\tau_t+\rho \nu u' =\rho f U^2 (1-y/R)
\end{equation} 
and
\begin{equation} \label{varep}
\varepsilon= f U^2 (1-y/R) u'-\nu u'^2.
\end{equation}

\begin{figure} 
\centering
\resizebox{2.2in}{!}{\includegraphics{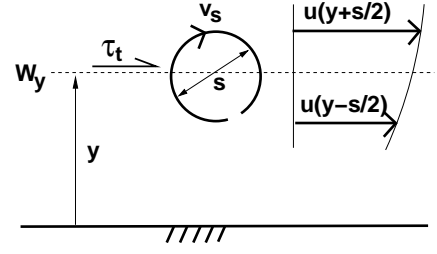}}
\caption{\label{mtrans} Schematic for the derivation
 of the turbulent shear stress. $u(y+s/2)$ is the mean
 velocity at a distance $y+s/2$ from the wall; 
$u(y-s/2)$ is the mean
 velocity at a distance $y-s/2$ from the wall. 
   }
\end{figure}

We now seek to derive an expression for
 the turbulent shear stress $\tau_t$.
 Let us call $W_y$ the wetted surface
at a distance $y$ from the wall of the pipe
 (Fig.~\ref{mtrans}).
The turbulent shear stress that acts on $W_y$
 is produced by eddies 
 that straddle $W_y$ and transfer momentum across $W_y$
 (Fig.~\ref{mtrans}).
Thus an eddy 
 of size $s$
 carries fluid of high horizontal momentum 
 per unit volume (about $\rho u(y+s/2)$)
 downward across $W_y$ and fluid of low horizontal
 momentum per unit volume (about $\rho u(y-s/2)$) upwards
  across $W_y$, and the eddy
 may be said to span a momentum contrast 
  $\rho (u(y+s/2)-u(y+s/2))\approx \rho s u'(y)$ \cite{firsterm}.
 The rate of momentum transfer across $W_y$ 
 is set by the velocity of the eddy, $v_s$
 (i.e., the velocity normal to $W_y$). 
 Therefore, the turbulent shear 
 stress produced by an eddy of size $s$ 
  scales as the product of
  the momentum contrast times the rate of momentum transfer, 
 or $\rho s u'(y)\, v_s$.

 In order to  identify the {\sl dominant\/} eddies 
 that straddle $W_y$ 
(i.e., the eddies that dominate the momentum 
transfer across $W_y$), 
recall that $v_s$
 (and therefore $s v_s$) increases with $s$.  
 It follows that the larger the eddy the 
 larger the turbulent shear stress 
 produced by the eddy.
  Nonetheless, eddies much larger than $y$ 
 do not properly straddle $W_y$ and
  can provide only a negligible velocity
 normal to $W_y$, and therefore a negligible 
 turbulent shear stress.
 (This observation is purely a matter of
  geometry.) We conclude that the dominant
 eddies that straddle $W_y$ are the eddies of
 size $s=y$. The turbulent shear stress 
 at a distance $y$ from the wall is thus given by
  $\tau_t=\kappa_\tau  \rho y v_y u'(y)$,
 where $\kappa_\tau$ is a dimensionless 
 proportionality constant.

 We are now ready to obtain an equation
 for the MVP. We substitute (\ref{varep})
 in $v_y= (\kappa_\varepsilon \varepsilon y)^{1/3}\sqrt{I}$,
 substitute the resulting expression for $v_y$ 
  in $\tau_t=\kappa_\tau  \rho y v_y u'$, 
and substitute the resulting expression for $\tau_t$
  in (\ref{tau}). After some algebra, 
 we write the final result in terms
 of ${\rm Re}=UR/\nu$ and the dimensionless variables 
 ${\hat y}\equiv y/R$ and ${\hat u}\equiv u/U$:
\begin{equation} \label{lau}
 \kappa^2 I^3 {\hat y}^2 {\hat u}'^{2}
 +{\rm Re}^{-1} {\hat u}'-f (1-{\hat y})=0,
\end{equation}
where $\kappa \equiv(\kappa_\varepsilon \kappa_\tau^3)^{1/4}$,
${\hat u}'\equiv d{\hat u}/d{\hat y}$ and 
 $I={\mathcal I}(\eta/y,{\hat y})$. To obtain 
an equation for 
 $\eta/y$  we substitute (\ref{varep}) in
 $\eta=\nu^{3/4}\varepsilon^{-1/4}$ and change 
 variables to $\hat y$ and $\hat u$:
\begin{equation} \label{laeoy}
 \eta/y={\rm Re}^{-1/2}\,(f {\rm Re}\, {\hat u}' (1-{\hat y})- {\hat u}'^2)^{-1/4}\,{\hat y}^{-1}.
\end{equation}

 
 If for a fixed Re we let ${\hat y}\rightarrow 0$,
 then $\eta/y\rightarrow \infty$ (from (\ref{laeoy})),
  $I \rightarrow {\mathcal I}(\infty,{\hat y})=0$, 
 and (\ref{lau}) simplifies to 
  ${\hat u}'=f {\rm Re}$, which is the law of the viscous layer.
  If for a fixed ${\hat y}\ll 1$ we let ${\rm Re}\rightarrow \infty$,
 then $\eta/y \rightarrow 0$ (from (\ref{laeoy})),
  $y$ is in the inertial range (where $I=1$), and 
 (\ref{lau}) simplifies to
 ${\hat u}'= \sqrt{f} / \kappa {\hat y}$,
 which we recognize as the log-law of the wall with a K\'arm\'an 
 constant $\kappa=(\kappa_\varepsilon \kappa_\tau^3)^{1/4}$. 
Note that 
the log law of the wall prevails where $y$ is in the inertial 
range; it follows that the dominant eddies in the log layer 
are eddies of the inertial range. 
 The presence of $\kappa_\varepsilon$ and $\kappa_\tau$
 in the expression, $\kappa=(\kappa_\varepsilon \kappa_\tau^3)^{1/4}$,
 reminds us of the underpinnings of the theory: the spectrum and 
 the momentum transfer, respectively. 
As $\kappa_\varepsilon$ is fixed by 
Kolmog\'orov's four-fifth law, which gives $\kappa_\varepsilon$ = 4/5 
 \cite{ff,ff1},  the K\'arm\'an constant
$\kappa$ is but an alternative form of the momentum-transfer 
constant $\kappa_\tau$. 
 From  $\kappa$= 0.42 (the experimental value \cite{exper}), 
we have $\kappa_\tau$ = 0.34. 




  The law of the viscous layer
  and the the log law of the wall may be made 
  invariant to changes in Re and $f$, in the form
  ${\tilde u}'=1$ and ${\tilde u}'=1 / \kappa {\tilde y}$, 
 by choosing ${\tilde y}\equiv {\rm Re} \sqrt{f}\, {\hat y}
  = {\rm Re} \sqrt{f}\, y/R$ 
 and ${\tilde u}\equiv {\hat u}/\sqrt{f}=  u/U\sqrt{f}$, 
 which we recognize as the wall variables.
 In the wall variables (\ref{lau}) becomes
\begin{equation} \label{util}
\kappa^2 I^3 {\tilde y}^2 {\tilde u}'^{2}
 + {\tilde u}'- (1-{\tilde y}/ {\rm Re} \sqrt{f})=0,
\end{equation}
where $I\equiv {\mathcal I}(\eta/y,{\tilde y}/{\rm Re}\sqrt{f})$, 
 and (\ref{laeoy}) becomes
 \begin{equation} \label{etatil}
\eta/y= ( {\tilde u}' (1-{\tilde y}/{\rm Re}\sqrt{f})
 -{\tilde u}'^2)^{-1/4}\,{\tilde y}^{-1}.
\end{equation}
 From these equations it is apparent that in the wall
 variables there is a single MVP
 except close to the centerline of the pipe, 
 where ${\tilde y}\approx {\rm Re}\sqrt{f}$ (or $y\approx R$).

\begin{figure} 
\centering
\resizebox{3.5in}{!}{\includegraphics{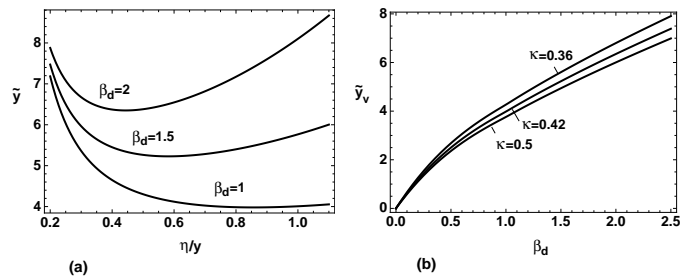}}
\caption{\label{yvisfig} (a) Plots of (\ref{ecuy}) for $\kappa=0.42$ and three
  values of $\beta_d$. (b) Plots of the thickness
 of the viscous layer ${\tilde y}_v$ as 
 a function of $\beta_d$ for three  values of $\kappa$.  }
\end{figure}

  We now ascertain under what conditions 
 (\ref{util}) and (\ref{etatil}) are compatible with a
  nonvanishing turbulent shear stress close to the wall.
 Suppose that at a point ${\tilde y}\ll {\rm Re}\sqrt{f}$ (or $y\ll R$)
   we have $I={\mathcal I}(\eta/y,0)>0$,
  and therefore $\tau_t>0$.
Then, we can eliminate ${\tilde u}'$ from (\ref{util}) and (\ref{etatil}) 
to obtain 
\begin{equation} \label{ecuy}
{\tilde y}=\left({ {(\eta/y)^{4/3} + \kappa^{4/3}\, {\mathcal I}^2(\eta/y,0) }
\over{\kappa^{2/3}\, (\eta/y)^{8/3}\, {\mathcal I}(\eta/y,0) }} \right)^{1/2}.
\end{equation}
 Plots of ${\tilde y}$ vs.\ $\eta/y$ (Fig.~\ref{yvisfig}) 
reveal that for any given $\kappa$ and $\beta_d$ there exists a 
  minimum value of ${\tilde y}$, to be denoted ${\tilde y}_v$. 
  Therefore,  for  ${\tilde y}<{\tilde y}_v$ it must be 
 that $I=0$, $\tau_t=0$, and ${\tilde u}'=1$.
 As this latter equation is the law of the viscous layer,
  we identify ${\tilde y}_v$ with the thickness of 
the viscous layer \cite{layers}. Note that ${\tilde y}_v$ depends on 
 the dissipative-range parameter $\beta_d$ 
 (Fig.~\ref{yvisfig}b), and
  that for $\beta_d=0$ there is no viscous layer (${\tilde y}_v=0$).

\begin{figure} 
\centering
\resizebox{3.4in}{!}{\includegraphics{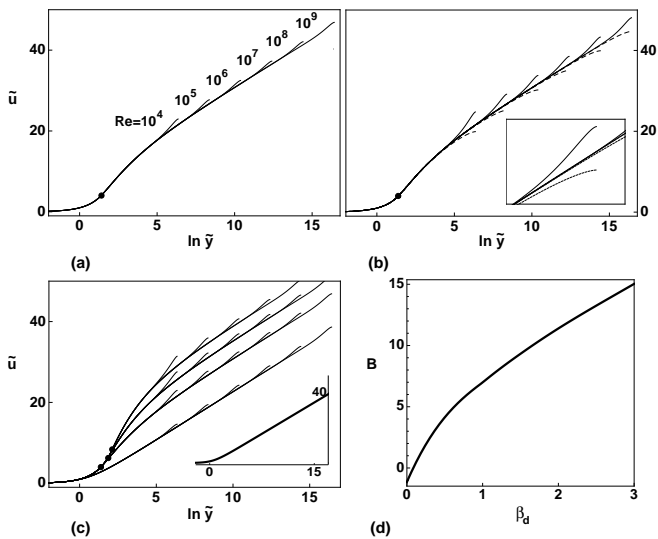}}
\caption{\label{ufig} (a) The MVPs computed using 
  $\kappa=0.42$, $\beta_e=3$, and $\beta_d=1$.
  The thick dot 
 indicates the point of contact between the viscous layer and 
 the buffer layer. (b) The same as in (a) but using 
  $\beta_e=0$ (dashed lines) and $\beta_e=7$ (solid lines).
Inset: detail of the wakes.
  (c) The same as in (a) but using 
   $\beta_d=0$ (bottom), 1, 2, and 3 (top).
 Inset: A plot of 
(\ref{ubeta0}) for $\kappa=0.42$.
(d) A plot of $B$ 
as a function $\beta_d$ for $\kappa=0.42$. }
\end{figure}

We are now ready to compute the MVP.
For simplicity we use 
  the Blasius relation for the friction factor,
 $f=0.033\,{\rm Re}^{-1/4}$ \cite{exper}. 
  Values of the spectral parameters
  $\beta_d$ and $\beta_e$ 
  from the fitting of 
  spectra are of O(1) \cite{pope};
  here we use $\beta_d=1$ and $\beta_e=4$. 
 The thickness of the viscous layer is set by 
 $\beta_d$ and $\kappa$;
 for $\beta_d=1$ and $\kappa=0.42$, ${\tilde y}_v= 3.9$
 (Fig.~\ref{yvisfig}b).  Therefore, for
  ${\tilde y}<3.9$ we can write ${\tilde u}({\tilde y})={\tilde y}$, 
 and for   ${\tilde y} >3.9$ we must compute ${\tilde u}({\tilde y})$
 by integrating (\ref{util}) with
 boundary condition ${\tilde u}(3.9)=3.9$.
The results are shown in Fig.~\ref{ufig}a.
 Given the spectrum,
 the theory yields the entire MVP with all of its distinctive 
 features. The specific conection between each one of 
 these features and the spectrum will become apparent 
 in what follows.

To elucidate the effect of the 
 energetic-range correction, we recompute the MVP
  using $\beta_e=7$ (a larger value than before)
 and $\beta_e=0$ (the smallest possible value, 
 which corresponds to having no energetic-range 
 correction). 
 The results (Fig.~\ref{ufig}b)
  indicate that 
 the energetic-range 
  correction steepens the MVP in the wake.
 The dominant eddies 
 in the wake must therefore
  be eddies of
 the energetic range. These eddies are slowed
 down by the energetic-range correction (and 
 made less adept at transfering momentum). 
 This effect explains the steepening
 of the MVP in the wake.
 
We have seen that the 
  dissipative-range correction sets
  the thickness of the viscous layer 
(Fig.~\ref{yvisfig}b)).
To understand further the effects of the 
 dissipative-range correction, we recompute the MVP
 using a few values of $\beta_d$, including
 $\beta_d=0$ (the smallest possible value, 
 which corresponds to having no dissipative-range 
 correction). 
 The results (Fig.~\ref{ufig}c) 
 indicate that the dissipative-range  
  correction causes the buffer layer to form, 
 so that for $\beta_d=0$ there is no buffer layer.
 (The buffer layer is the part of the MVP where 
 the MVP has a 
 negative curvature \cite{layers}.) 
The dominant eddies in the buffer layer 
must therefore be eddies of the dissipative range. 
These eddies
  are slowed down by the dissipative-range 
 correction; the larger the eddies, the 
  less they are slowed down, and the more adept 
 they remain at transferring momentum. As the size of the 
 dominant eddies increases with the distance
 to the wall, the MVP becomes less steep
 as we traverse the buffer layer from the 
  outer edge of the viscous layer 
 (where the eddies are fully viscous) 
 to the inner edge of
  the log layer (where the eddies are
  fully inertial).
  This effect explains
 the negative curvature of the MVP in the buffer layer.

It is apparent from Fig.~\ref{ufig}c that the constant $B$ of the
log law of the wall is set by $\beta_d$. 
A plot of $B$ as a function 
of $\beta_d$ is shown in Fig.~\ref{ufig}d.

 %


 Interestingly,
  the MVP for $\beta_d=0$ (no dissipative-range correction)
 can be obtained  analytically everywhere save the wake. In fact,
  for $\beta_d=0$ and 
 ${\tilde y}\ll {\rm Re}\sqrt{f}$ (or $y\ll R$),
 $I=1$ and (\ref{util}) simplifies to 
 $\kappa^2 {\tilde y}^2 {\tilde u}'^{2}+ {\tilde u}'- 1=0$,
 which with  ${\tilde u}(0)=0$ yields
\begin{equation} \label{ubeta0}
{\tilde u}={{ {\rm arcsinh}(2 \kappa {\tilde y})}\over{\kappa}}+ 
 {{1-\sqrt{1+\kappa^2 {\tilde y}^2 }\over{2 \kappa^2 {\tilde y}}}}.
\end{equation}
 This is what the MVP would be away from the wake 
 if the Kolmog\'orov spectrum
 were valid even for vanishingly small eddies (inset of Fig.~\ref{ufig}c).
For ${0\ll \tilde y}\ll {\rm Re}\sqrt{f}$ (or $0\ll y\ll R$),
 (\ref{ubeta0}) simplifies asymptotically to  
${\tilde u}\sim (1/\kappa) \ln {\tilde y} + B$, with 
$B = (-1+\ln 4\kappa)/\kappa$.
Thus, for 
$\beta_d=0$ and $\kappa=0.42$, $B=-1.15$ 
in accord with Fig.~\ref{ufig}d.

We have established the long-surmised link \cite{tl} between
the mean velocity profile and the turbulent spectrum. To
test our results, we have shown that the usual model of
the spectrum (a power-law inertial range with corrections
for the dissipative range and the energetic range)
is in itself sufficient to compute with no additional assumptions
a mean velocity profile complete with viscous
layer, buffer layer, log layer, and wake. The thickness
of the viscous layer, the two constants of the log law of
the wall, and the amplitude of the wake are all set by
the dimensionless momentum-transfer constant $\kappa_\tau$ and
the usual spectral parameters---the parameter $\beta_e$ of the
energetic-range correction and the parameter $\beta_d$ of the
dissipative-range correction. The relation between a specific
feature of the MVP and the spectral parameters reminds
us of the underlying physics. Thus, for example,
the K\'arm\'an constant is independent of both $\beta_e$ and $\beta_d$,
and is therefore unaffected by the energetic-range and
dissipative-range corrections, with the implication that
the eddies that dominate the momentum transfer in the
log layer are eddies of the inertial range. More broadly,
the close relation between the mean velocity profile and
the turbulent spectrum indicates that in turbulence, as
in continuous phase transitions, global variables are governed by the
statistics of the fluctuations.

\begin{acknowledgments}
 NSF funded this work through
  grant DMR06--04435. 
\end{acknowledgments}


\begin{thebibliography}{}
\bibitem{layers} We define the viscous layer, the 
   buffer layer, etc.\ (caption of Fig.~\ref{expe}),
as in ch.\ 5 of \cite{tl}, in particular Fig. 5.6.
\bibitem{ff}The four-fifth law reads
 $\overline{v_s^3} = -(4/5)\, \varepsilon \,s$,
 where the left-hand side
 is the third-order structure function.
 The estimate $|\overline{v_s^3}|=u_s^3$ leads to
 $\varepsilon=(5/4)\, u_s^3/s$, or $\kappa_\varepsilon=4/5$.
\bibitem{firsterm} We retain only the 1st term of the Taylor
  series because (a) additional terms 
 would necessitate BCs other than 
 the no-slip BC and (b) in our final solution for $u(y)$  
the 2nd term $\le 11$\% of the 1st term for all $y$.
\end{thebibliography}

\end{document}